\documentclass[preprint,prd,noshowpacs,nofootinbib]{revtex4}
\usepackage{amssymb}
\usepackage{amsmath}

\def\be {\begin{equation}}
\def\ee {\end{equation}}
\def\bea {\begin{eqnarray}}
\def\eea {\end{eqnarray}}

\begin{document}
\title{{\bf Modified Dispersion Relation, Photon's Velocity,\\ and Unruh Effect}}

\author{Bibhas Ranjan Majhi $^1$} \email[email: ]{bibhas@iucaa.ernet.in}
\author{Elias C. Vagenas $^2$} \email[email: ]{evagenas@academyofathens.gr}

\affiliation{$^1$~IUCAA, Post Bag 4, Ganeshkhind,
Pune University Campus, Pune 411 007, India \\}

\affiliation{$^2$~Research Center for Astronomy and Applied
Mathematics, Academy of Athens, Soranou Efessiou 4, GR-11527,
Athens, Greece}

\begin{abstract}
\par\noindent
Motivated by the recently derived new form of generalized uncertainty principle we obtain the corresponding dispersion
relation which is now modified. This modification can be interpreted as a possible mechanism that makes particles more massive.
In addition, the modified velocity of photons is obtained and indicates that photons' propagation depends on their energy, thus
superluminal photons are permitted due to generalized uncertainty principle. Furthermore, we derive and solve the two-dimensional
Klein-Gordon equation in the presence of GUP corrections, and therefore the GUP-corrected emission spectrum due to the Unruh effect is obtained.
\end{abstract}

\maketitle

\section{Introduction}
It was in the sixties when for the first time the concept of fundamental measurable length due to the effect of gravity
on  experiments was introduced \cite{Mead:1964zz}. This minimum measurable length gave rise to the modification of Heisenberg uncertainty principle,
nowadays known as Generalized Uncertainty Principle (GUP) \cite{guppapers}. Recently, a new form of GUP consistent with  String Theory, Black Hole Physics, and DSR,
was introduced which implies not only a minimum measurable length but also a maximum measurable momentum \cite{Ali:2009zq}
\footnote{Similar commutators and forms of generalized uncertainty principle with those in \cite{Ali:2009zq} were proposed and derived
in \cite{Kempf:1994su,Kempf:1996fz,Brau:1999uv,Das:2008kaa}.}.
Several phenomenological implications of this new form of GUP have been provided by the same authors and their collaborators \cite{Das:2009hs,Das:2010zf}.

In this letter, we employ this new form of GUP in order to obtain the corresponding dispersion relation and photons' velocity.
In particular, in section II we derive the modified dispersion relation that corresponds to new form of GUP and in section III we
obtain the modified velocities of photons which are energy-dependent. In section IV, we write and solve the two-dimensional
Klein-Gordon equation which is now modified due to GUP corrections. In section V, using the positive frequency mode solution of the modified
Klein-Gordon equation we derive the emission spectrum of Unruh effect. The spectrum is GUP-corrected and in the limiting case in which the
corrections are eliminated the standard emission spectrum is obtained. Finally, in section VI  results  are briefly discussed.

\section{Modified Dispersion Relation}
In this section we will derive the dispersion relation that corresponds to the new form of GUP as introduced in \cite{Ali:2009zq}.
Let us denote with $p^a$  the modified four momentum (or, equivalently,  the momentum at high energies)
while with $k^a$  the usual four momentum (or, equivalently,  the momentum at low energies which satisfies the standard representation in position space).
 The modified and the usual momenta are related by the following relations
\begin{eqnarray}
&&p^0 = k^0;
\label{1.00}
\\
&&p^{i} = k^{i}(1-\alpha {\bf{k}} + 2\alpha^2{\bf{k}}^2)~,
\label{1.01}
\end{eqnarray}
where $\alpha  = \alpha_{0} / M_{Pl} c$ is a small parameter, $M_{Pl}$ is the Planck mass, and ${\bf{k}}^2 = g_{ij} k^{i}  k^{j}$, thus ${\bf{k}} = \sqrt{g_{ij} k^{i}  k^{j}}$.
In our convention, the capital Latin indices, such as $A,B,\ldots,$ take the values  $0,1,2,3,.....$
while the spatial indices are denoted by the small Latin indices $i,j,\ldots = 1,2,3,....$.
\par\noindent
We now consider the gravitational background metric
\begin{eqnarray}
ds^2 =g_{AB}dx^{A}dx^{B} =  g_{00} c^{2} dt^2 + g_{ij} dx^{i} dx^{j}
\label{1.02}
\end{eqnarray}

and the square of the four-momentum is this background reads
\begin{eqnarray}
p^{A}p_{A} &=& g_{AB}p^{A}p^{B} = g_{00}(p^0)^2 + g_{ij}p^{i} p^{j}
\nonumber
\\
&=& g_{00}(k^0)^2 + g_{ij} k^{i} k^{j} (1-\alpha {\bf{k}} + 2\alpha^2{\bf{k}}^2)^2~.
\label{1.03}
\end{eqnarray}
Keeping terms up to ${\cal{O}}(\alpha^2)$ in the above expression we obtain
\begin{eqnarray}
p^{A} p_{A} =  g_{00}(k^0)^2 + {\bf k}^2 + {\bf k}^2 (-2\alpha{\bf k} + 5\alpha^2{\bf k}^2)~.
\label{1.04}
\end{eqnarray}
It is evident that the first two terms in the RHS of equation (\ref{1.04}) form
the usual dispersion relation $k^{A}k_{A} = g_{00}(k^0)^2 + {\bf k}^2 = -m^{2} c^{2}$.
Thus, equation (\ref{1.04}) reads now
\begin{eqnarray}
p^{A} p_{A} = -m^{2}c^{2} + {\bf k}^2 (-2\alpha{\bf k} + 5\alpha^2{\bf k}^2)~.
\label{1.05}
\end{eqnarray}
Inverting equation (\ref{1.01}) in order to express the usual momentum in terms of the modified one,
and keeping terms up to ${\cal{O}}(\alpha)$, the usual momentum reads
\begin{eqnarray}
k^{i} = p^{i} (1+\alpha {\bf p})~.
\label{1.06}
\end{eqnarray}
Substituting the above expression for the usual momentum and retaining terms up to ${\cal{O}}(\alpha)$,
the modified dispersion relation that corresponds to the new form of GUP \cite{Ali:2009zq} takes the form
\begin{eqnarray}
p^{A}p_{A} = -m^{2} c^{2} -2\alpha{\bf p}{\bf p}^2~. \label{1.07}
\end{eqnarray}
At this point, a couple of comments are in order.
First, it is easily seen from the modified equation, i.e. equation (\ref{1.07}), that the RHS of this equations can be viewed as a modified mass
due to the quantum gravity corrections. Thus, the quantity $\sqrt{m^{2} + 2\alpha{\bf p}{\bf p}^2 / c^{2}}$ can be interpreted as an effective mass.
Second, it is known that the energy of a particle is defined through the expression $\frac{E}{c} = - \xi_{A} p^{A} = - g_{AB}\xi^{A} p^{B}$,
where $\xi^{A}$ is the Killing vector. Setting $\xi^{A} = (1,0,0,....)$, then the energy in the gravitational background with metric
(\ref{1.02}) is given as $E = - g_{00} c p^0$.
Using equation (\ref{1.07}), the time component of the momentum, i.e. $p^0$, can be written as
\begin{eqnarray}
(p^0)^2 = \frac{1}{g_{00}}[-m^2 c^{2} - {\bf p}^2 (1+2\alpha{\bf p})]~.
\label{1.08}
\end{eqnarray}
Therefore, the energy of a particle can be expressed in terms of the mass, $m$,
and of the three spatial momentum ${\bf p}$ as follows
\begin{eqnarray}
E^2 = (- g_{00} c p^0)^2 = - g_{00}[m^{2} c^{4} + c^{2} {\bf p}^{2}  (1+2\alpha{\bf p})]~.
\label{1.09}
\end{eqnarray}
For the special case of Minkowski spacetime in which  $g_{00} =
-1$ and in absence of quantum gravity corrections, i.e. $\alpha
=0$, we retrieve the standard dispersion relation $E^2 = m^{2}
c^{4} + c^{2} {\bf k}^2$.
At this point it is worth mentioning that our result, i.e. the modified dispersion relation in equation (\ref{1.09}), is similar to
other modified dispersion relations already existing in the literature (for instance, see \cite{AmelinoCamelia:2008qg} and references therein).
However, this is the first time to our knowledge that the modified dispersion relation of the new form of GUP  proposed by Ali, Das and Vagenas \cite{Ali:2009zq}
is obtained.

\section{Photon's Velocity}

In this section we will obtain the modified velocities of photons for the specific gravitational background given by equation  (\ref{1.02}).
 The velocity of a photon is defined as $u = \frac{1}{\sqrt{-g_{00}}}\frac{\partial E}{\partial{\bf p}}$.
 Setting  $m=0$  in equation (\ref{1.09}), we obtain for photons
\begin{eqnarray}
\frac{\partial E}{\partial{\bf p}} &=& \pm (-g_{00})^{\frac{1}{2}} c (1+3\alpha {\bf p}) (1+2\alpha {\bf p})^{-\frac{1}{2}}
\nonumber
\\
&\simeq& \pm (-g_{00})^{\frac{1}{2}} c (1+2\alpha {\bf p})\qquad [\mbox{up to ${\cal{O}}(\alpha)$}]~.
\label{1.10}
\end{eqnarray}
In order to express the above expression in terms of energy $E$,
equation (\ref{1.09}) has to be solved  for ${\bf p}$. This can be
done by iteration method. Thus, the zeroth order solution for
${\bf p}$ is obtained when  $\alpha$ is set equal to zero in
equation (\ref{1.09})
\begin{eqnarray}
{\bf p} = \pm \frac{1}{\sqrt{-g_{00}}}\frac{E}{c}~.
\label{1.11}
\end{eqnarray}
Next, we substitute back the zeroth order solution for ${\bf p}$, namely equation (\ref{1.11}),
in equation (\ref{1.09}) and we get the solution of ${\bf p}$ in terms of $E$
\begin{eqnarray}
{\bf p} = \pm \frac{E}{c \sqrt{-g_{00}}}\Big(1\mp \frac{\alpha E}{c \sqrt{-g_{00}}}\Big)~.
\label{1.12}
\end{eqnarray}
Substituting equation (\ref{1.12}) in equation (\ref{1.10}) and keeping terms up to ${\cal{O}}(\alpha)$, we get
\begin{eqnarray}
\frac{\partial E}{\partial{\bf p}} = \pm (-g_{00})^{\frac{1}{2}} c (1\pm \frac{2\alpha E}{c \sqrt{-g_{00}}})~,
\label{1.13}
\end{eqnarray}
and thus velocity of the photon reads
\begin{eqnarray}
u = \frac{1}{\sqrt{-g_{00}}}\frac{\partial E}{\partial{\bf p}} =
c  \Big(\pm 1 + \frac{2\alpha E}{c \sqrt{-g_{00}}}\Big)~.
\label{1.14}
\end{eqnarray}
At this point, a number of comments are in order.
First, if one considers the special case of Minkowski spacetime, i.e. $g_{00} = -1$, with no quantum gravity corrections, namely $\alpha = 0$,
one should obtain from equation (\ref{1.14}) the velocity of photon to be constant and equal to $c$. So, in the above expression for the photon velocity,
we are eligible to discard the negative sign and consider only the positive one. Therefore, the modified photon velocity turns out to be
\begin{eqnarray}
u = c \Big(1 + \frac{2\alpha E}{c \sqrt{-g_{00}}}\Big)~.
\label{1.15}
\end{eqnarray}
Second, it is evident that there is a velocity dispersion and that the photon velocities are energy dependent. Moreover,
since photon velocity $u$ can take larger values than $c$, superluminal photon propagation is allowed due to quantum gravity corrections.
Third, our result given by equation (\ref{1.15}) is similar to the earlier result obtained in \cite{AmelinoCamelia:1997gz} (see their equation (1)).
Actually, if we demand these two results to be identical, then we can specify the arbitrary parameters of \cite{AmelinoCamelia:1997gz} as
\begin{eqnarray}
\xi =-1; \,\,\,\ E_{QG} = \frac{c \sqrt{-g_{00}}}{2\alpha_{0}} E_{Pl}~.
\label{1.16}
\end{eqnarray}
%
%
%
%
\section{Modified Klein-Gordon equation in 2D Minkowski Space}
Our goal in this section is to obtain and solve the GUP-corrected Klein-Gordon equation in two dimensions.
The line element of Minkowski spacetime in two-dimensions is of the form
\begin{eqnarray}
ds^2 = -dT^2 + dX^2~.
\label{minkowski}
\end{eqnarray}
Comparing equation (\ref{minkowski}) with equation (\ref{1.02}), the gravitational background under study will now have
$g_{00} = g_{TT} = -1$, $g_{11}= g_{XX} = +1$ and hence $k^{A} = (k^{T},k^{X})$.
Therefore, using the standard operator position representation
\begin{eqnarray}
k^{0}= k^{T} =  i\hbar\frac{\partial}{\partial T}; \,\,\ k^{1}=k^{X} = - i\hbar \frac{\partial}{\partial X}
\label{1.17}
\end{eqnarray}
in equation (\ref{1.04}) and keeping terms up to ${\cal{O}}(\alpha)$, the modified Klein-Gordon equation reads
\begin{eqnarray}
p^{A} p_{A}\Phi(T,X) =  \Big[\frac{\partial^2}{\partial T^2} - \frac{\partial^2}{\partial X^2} - 2i\alpha\hbar\frac{\partial^3}{\partial X^3}\Big]\Phi(T,X) = 0~.
\label{1.18}
\end{eqnarray}
It is evident that the first two terms are the standard ones while the third term is due to the GUP corrections (see also \cite{Das:2010zf}).
\par\noindent
We are now interested in getting positive frequency mode solutions of equation (\ref{1.18}) and for this we choose a solution of the form
\begin{eqnarray}
\Phi(T,X) = e^{-i\omega T}\Psi(X)~.
\label{1.19}
\end{eqnarray}
Substituting the aforesaid solution in equation (\ref{1.18}) we obtain
\begin{eqnarray}
\frac{d^2 \Psi(X)}{d X^2} + 2i\alpha\hbar\frac{d^3 \Psi(X)}{d X^3} + \omega^2\Psi(X) = 0
\label{1.20}
\end{eqnarray}
and employing the ansatz $\Psi(X) = e^{nX}$, we obtain an equation for $n$
\begin{eqnarray}
n^2 + \omega^2 + 2i\alpha \hbar n^3 = 0~.
\label{1.21}
\end{eqnarray}
To find the roots of  equation (\ref{1.21}), the iteration method will be utilized and explicit expressions of
them will be given up to the leading order in $\alpha$ since all previous expressions are also
written by keeping terms up to the leading order in $\alpha$. Within this order, the roots are of the form (see also \cite{Ali:2009zq})
\begin{eqnarray}
n_1 = i\omega(1+\alpha\hbar\omega); \,\,\ n_2 = -i\omega (1-\alpha\hbar\omega);
 \,\,\,\ n_3 = \frac{i}{2\alpha\hbar} - 2i\alpha\hbar\omega^2 \simeq \frac{i}{2\alpha\hbar} ~.
\label{1.22}
\end{eqnarray}
For later purpose, we will concentrate on the outgoing solution. The main reason is that the ingoing modes are
actually trapped inside the Rindler horizon while the outgoing modes can escape and  this can be viewed as the radiation emanating from the horizon.
Therefore, in order to discuss about the radiation spectrum, as perceived by the uniformly accelerated observer, it is important to
concentrate solely on the outgoing modes. The second solution $n_2$ corresponds to the ingoing solution while $n_3$ is independent of $\omega$.
Therefore, for the present study, these two solutions are considered as unimportant. The first solution $n_1$ is the only solution that leads to positive spatial momentum,
and thus it is the only that will be considered in the subsequent analysis. The positive frequency outgoing solution of the modified Klein-Gordon equation now reads
\begin{eqnarray}
\Phi(T,X) = e^{-i\omega(T-X) + i\alpha\hbar\omega^2 X}~.
\label{1.23}
\end{eqnarray}
At this point, it should be stressed that equation (\ref{1.23}) is  a solution of Klein-Gordon equation in two-dimensional
Minkowski spacetime, valid up to ${\cal{O}}(\alpha)$, and under the condition that it represents the  positive frequency outgoing mode solution.
%
%
%
%
\section{Unruh effect}
It is known that Unruh effect is a phenomenon that emerges due to the ``marriage" of General Relativity and Quantum Field Theory.
We expect that  the emission spectrum related to the Unruh effect will be modified due to GUP 
 \cite{Rinaldi:2008qt,Gutti:2010nv,Harikumar:2012yu,Harikumar:2012ff}.  
Therefore, in this section we will discuss the Unruh effect in the two-dimensional Minkowski spacetime and derive its emission spectrum.

Let us consider a uniformly accelerated frame, which is usually called as Rindler frame. The coordinates transformations
from Minkowski to Rindler frame with respect to a Rindler observer moving along the $x$-axis are
\begin{eqnarray}
X(\tau) = \frac{1}{\kappa}{\textrm{cosh}}(\kappa\tau); \,\,\,\ T(\tau) = \frac{1}{\kappa}{\textrm{sinh}}(\kappa\tau)~.
\label{1.24}
\end{eqnarray}
The accelerated observer will now observe the wave function (\ref{1.23}) in the form
\begin{eqnarray}
\Phi\Big[T(\tau),X(\tau)\Big]\equiv \Phi[\tau] = {\textrm {exp}}\Big[{\frac{i\omega}{\kappa}e^{-\kappa\tau}(1+\frac{\alpha\hbar\omega}{2})
+\frac{i\alpha\hbar\omega^2}{2\kappa}e^{\kappa\tau}}\Big]
\label{1.25}
\end{eqnarray}
and thus the modified mode frequency that the accelerated observer will detect will be
\begin{eqnarray}
\Omega(\tau) = \omega e^{-\kappa\tau} + \frac{\alpha\hbar\omega^2}{2}(e^{-\kappa\tau}+e^{\kappa\tau})~.
\label{1.26}
\end{eqnarray}
%
%
%
%
%
\par\noindent
The power spectrum of the mode (\ref{1.25}) will be given by the relation $P(\nu) = |f(\nu)|^2$ with  $f(\nu)$ to be the Fourier transform of $\Phi(\tau)$
with respect to $\tau$
\begin{eqnarray}
\Phi(\tau) = \int_{-\infty}^{+\infty}\frac{d\nu}{2\pi}f(\nu)e^{-i\nu\tau}~.
\label{1.27}
\end{eqnarray}
Employing equation (\ref{1.25}), the Fourier transform of  $\Phi(\tau)$ reads
\begin{eqnarray}
f(\nu) = \int_{-\infty}^{+\infty}d\tau~\Phi(\tau)e^{i\nu\tau}
= \int_{-\infty}^{+\infty}d\tau~{\textrm{exp}}\Big[{i\nu\tau} + \frac{i\omega}{\kappa}(1+\frac{\alpha\hbar\omega}{2})e^{-\kappa\tau} +
\frac{i\alpha\hbar\omega^2}{2\kappa}e^{\kappa\tau}\Big]~.
\label{1.28}
\end{eqnarray}
To perform the integration, we make a change of variables $u= e^{-\kappa\tau}$ and expand the last term of the exponential
so as to compute the spectrum by utilizing  the analysis in \cite{Padmanabhan:2010zzb}. Thus, retaining terms up to ${\cal{O}}(\alpha)$,
equation (\ref{1.28}) becomes
\begin{eqnarray}
f(\nu) &\simeq& \frac{1}{\kappa}\int_{0}^{+\infty}du~ (u)^{-\frac{i\nu}{\kappa} - 1}{\textrm{exp}}\Big[\frac{i\omega}{\kappa}(1+\frac{\alpha\hbar\omega}{2})u\Big]
\nonumber
\\
&+& \frac{i\alpha\hbar\omega^2}{\kappa^2}\int_{0}^{+\infty}du ~ u^{-\frac{i\nu}{\kappa} - 2}
{\textrm{exp}}\Big[\frac{i\omega}{\kappa}(1+\frac{\alpha\hbar\omega}{2})u\Big]~.
\label{1.30}
\end{eqnarray}
Following the analysis for the computation of the power spectrum as given in \cite{Padmanabhan:2010zzb} and
keeping terms up to ${\cal{O}}(\alpha)$,  the power spectrum of the mode (\ref{1.25}) is of the form
\begin{eqnarray}
|f(\nu)|^2 = \frac{2\pi}{\kappa\nu}\frac{1}{e^{\frac{2\pi\nu}{\kappa}} - 1}\Big[1-\frac{2\alpha\hbar\omega^3}
{\kappa^3(\frac{\nu^2}{\kappa^2} + 1)}\Big]
\label{1.36}
\end{eqnarray}
 while the corresponding power spectrum per logarithmic band in frequency is written as
\begin{eqnarray}
\nu |f(\nu)|^2 = \frac{2\pi}{\kappa}\frac{1}{e^{\frac{2\pi\nu}{\kappa}} - 1}\Big[1-\frac{2\alpha\hbar\omega^3}
{\kappa^3(\frac{\nu^2}{\kappa^2} + 1)}\Big]~.
\label{1.37}
\end{eqnarray}
At this point, a couple of comments are in order. First, it is obvious that the last term in equation (\ref{1.37})
includes all corrections due to GUP  while by taking the limit $\alpha\rightarrow 0$,  only the first term survives and
is equal to the Planckian power spectrum \cite{Padmanabhan:2010zzb}. Second, it should be noted that the spectrum $P(\nu)$, as given in equation (\ref{1.36}),
must be positive definite and so one needs to have
\begin{eqnarray}
2\alpha \hbar \Big(\frac{\omega}{\kappa}\Big)^3 < \Big(\frac{\nu^2}{\kappa^2}+1\Big)
\label{1.38}
\end{eqnarray}
which actually gives a bound on the maximum value of frequency $\omega$.
This is related to the fact that the GUP employed here introduces a minimal length as well as a maximum measurable momentum.
%
%
%
%
\section{Conclusions}
In this paper we have managed to derive the modified dispersion relation of the newly formed generalized uncertainty principle \cite{Ali:2009zq}.
This modification can be interpreted as an effective particle's mass and, in particular, it seems that GUP works as a mechanism
that makes particles more massive. In the limiting case of Minkowski spacetime and with no quantum gravity corrections to be present,
the standard dispersion relation is obtained. Additionally, it is shown that the velocities of photons are energy dependent
due to the generalized uncertainty principle and superluminal photon propagation is allowed. Though superluminality sounds quite unphysical,
since in Special Relativity photons travel at the speed of light in vacuum, one can make the reasonable assumption that the principle of
relativity no longer holds near the Planck scale $E_{Pl}$ \cite{Hossenfelder:2009nu}. Moreover, it was recently shown that superluminality
can be accommodated as an apparent phenomenon in General Relativity. In particular, it was shown that photons (which are expected to travel at the speed of light $c$)
can be subluminal or superluminal depending on the path they follow in the gravitational field as well as on the
position of the observer \cite{Lust:2011fx}. The verification, or not, of the afore-said statements is expected to be given mainly by astrophysical observations.
However, the authors expect that when all kinds of quantum gravity corrections are taken into consideration,
meaning a consistent theory of quantum gravity is obtained, superluminality will be forbidden and causality will be recovered.
Furthermore, the  two-dimensional Klein-Gordon equation modified due to GUP effects is obtained and solved. By selecting only its positive frequency mode solutions,
we derive the emission spectrum due to the Unruh effect. The emission spectrum is GUP-corrected and an upper bound to the value of
the mode frequency is imposed. This condition is probably related to the fact that the specific form of GUP utilized in the present paper introduces,
apart from the minimum observable length, a maximum measurable momentum.
\par\noindent
Finally, we would like to mention some theoretical implications of the modified dispersion relation. It is known that the mass of
a Schwarzschild black hole decreases continuously  during the evaporation process and, consequently, the Hawking temperature diverges.
This scenario is considered as ambiguous due to the fact that at the end of Hawking radiation, Planck scale physics play an important
role and till now we do not have in our hands such a theory, i.e. a theory that includes the Planck scale effects and thus describes the aforesaid scenario.
The modified dispersion relation, derived here, does incorporate the Planck scale effects and hence one can expect that it has some theoretical
importance in the black hole paradigm. For example, one can ask what would be the modified Schwarzschild spacetime. In this direction, a proposal has been given
earlier in literature, widely known as {\it rainbow gravity theory} \cite{Magueijo:2002xx}. As a consequence, it is expected to have some effects
on the thermodynamics of black holes. Apart from this, one may be interested to know the Planck scale effects in other branches of physics.
All these suggest that the modified dispersion relation can play a important role in physics.
%
%
%
%
%
%
%
\section{Acknowledgments}
We would like to thank D. Singleton for useful discussions and enlightening comments.

\end{document}